\documentclass{sig-alternate-05-2015}
\usepackage{amsmath,mathtools,mathptmx} 
\usepackage{tikz, tikzscale, pgf, pgfplots, pgflibraryarrows}
\usepackage[square,sort,comma,numbers]{natbib}
\usetikzlibrary{shapes.geometric}
\usepackage{url}
\usetikzlibrary{positioning}
\usepackage{subcaption, float}
\usepackage{comment}

\renewcommand{\thesubsubsection}{\thesubsection.\alph{subsubsection}}
\newfont{\mycrnotice}{ptmr8t at 7pt}
\newfont{\myconfname}{ptmri8t at 7pt}
\let\crnotice\mycrnotice%
\let\confname\myconfname%

\CopyrightYear{2016} 
\setcopyright{acmlicensed}
\conferenceinfo{DLRS '16,}{September 15 2016, Boston, MA, USA}
\isbn{978-1-4503-4795-2/16/09}\acmPrice{\$15.00}
\doi{http://dx.doi.org/10.1145/2988450.2988457}

\begin{document}

\title{Exploring Deep Space: Learning Personalized Ranking in a Semantic Space}

\numberofauthors{3}
\author{
\alignauthor
Jeroen B. P. Vuurens \\
       \affaddr{The Hague \\ University of Applied Science \\ Delft University of Technology} \\
       \affaddr{The Netherlands} \\
       \email{j.b.p.vuurens@tudelft.nl}
\alignauthor
Martha Larson \\
       \affaddr{Delft University of Technology}\\
       \affaddr{Radboud University Nijmegen}\\
       \affaddr{The Netherlands}\\
       \email{m.a.larson@tudelft.nl}
\alignauthor
Arjen P. de Vries \\
       \affaddr{Radboud University Nijmegen}\\
       \affaddr{The Netherlands}\\
       \email{arjen@acm.org}
}

\maketitle
\begin{abstract}
Recommender systems leverage both content and user interactions to generate recommendations that fit users' preferences. The recent surge of interest in deep learning presents new opportunities for exploiting these two sources of information. To recommend items we propose to first learn a user-independent high-dimensional semantic space in which items are positioned according to their substitutability, and then learn a user-specific transformation function to transform this space into a ranking according to the user's past preferences. An advantage of the proposed architecture is that it can be used to effectively recommend items using either content that describes the items or user-item ratings. We show that this approach significantly outperforms state-of-the-art recommender systems on the MovieLens 1M dataset.

\end{abstract}


 
\section{Introduction}


State-of-the-art collaborative-filtering systems recommend items by analyzing the history of user-item preferences. Alternatively, content-based systems analyze data about the items, and suggest items to a user that are most similar to the items she liked in the past. Past research has shown collaborative filtering to be more effective than content-based systems, however, it also has a few disadvantages over content-based models. Firstly, collaborative filtering requires a large quantity of user data to infer preference patterns between users. Secondly, these algorithms are generally considered less capable of recommending novel items, while novel items may be preferable over popular items for instance when a recommender system is repeatedly used to look for a job or a house \cite{fleder2009blockbuster, lops2011content}. In cases when collaborative filtering is less applicable, content-based approaches can be used to complement the list of recommendations.

In recent years we have seen a rise in the use of semantic space models for various tasks such as translation and analogical reasoning \cite{lecun2015deep}. In such a space, each element is represented as an abstract vector, which typically captures semantic properties of the elements and semantic relations between elements. 

In this work, we present a novel approach for the recommendation of items, that first structures items in a semantic space and then for a given user learns a function to transform this space into a ranked list of recommendations that matches the user's preferences. We show that the same architecture can be used to effectively recommend items using either the text of user reviews or user-item ratings. We evaluate this approach using the MovieLens 1M dataset, and show that the proposed approach using user-item ratings significantly outperforms state-of-the-art recommender systems.


\section{Semantic spaces for RecSys}

\subsection{Semantic spaces}

Lowe \cite{lowe2001towards} defines a semantic space model as a way of representing the similarities between contexts in a Euclidean space. A semantic space represents the \textbf{intersubstitutability} of items in context, i.e.\ items may effectively be substituted by nearby items in a semantic space. This definition is based on Firth's observervation that ``you shall know a word by the company it keeps'' \cite{firth1957synopsis}. The intuition for this distributional characterization of semantics is that whatever makes words similar or dissimilar in meaning, it must show up distributionally in the lexical company of the words.

When comparing highly-dimensional objects such as text documents, similarity measures are only reliable for nearly identical objects, since the ``curse of dimensionality'' makes dissimilar items appear equi-distant \cite{beyer1999nearest, bengio2003neural}. In a semantic space, the curse of dimensionality can be counteracted by representing items using non-sparse vector elements that describe the strength of the association with item-related data. Various methods have been proposed to learn semantic representations. Landauer and Dumais \cite{landauer1997solution} perform a Latent Semantic Analysis by considering the informativeness of words in documents, i.e.\ word co-occurrences that are evenly distributed over documents are less informative than those that are concentrated in a small subset. Lowe and McDonald \cite{lowe2000direct} used a log-odds-ratio measure to explicitly factor out chance co-occurrences.

\subsection{Towards recommendations}
\label{sec:theorydeepranking}

In this work, we propose to learn semantic item representations, for the task of recommending items to a user. The key idea is to position all items in a high-dimensional normalized semantic space, in such a way that items that are more likely to substitute each other are positioned closely together. Ideally, the items are positioned in such a way that for each user there is a region that exclusively contains items that the user (knowing or unknowingly) likes, making it possible to recommend items to a user by simply finding the best region in semantic space. The substitutability between items can be inferred from the observation of being jointly liked by a subset of users, or in a content-based setting by having similar descriptions. 

\begin{figure}[!tbp]
  \centering
       \includegraphics[]{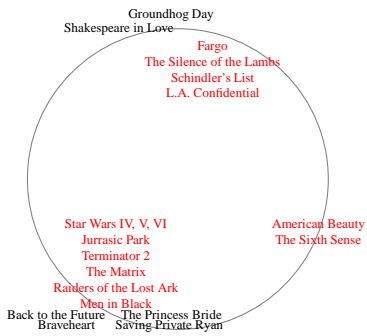}
   \caption{Example of a semantic space for the 20 most popular movies in MovieLens 1M. The figure is a normalized 2D t-SNE projection of the MovieLens user-item matrix. In red are movies that are positioned very closely and therefore represented as a cluster.}
  \label{fig:semanticspace}
\end{figure}

To illustrate such a semantic space, Figure~\ref{fig:semanticspace} shows a normalized t-SNE projection for movies in the MovieLens 1M dataset, representing every movie as a vector over the ratings by users. Using 2 dimensions, such a normalized space is shaped like the edge of a circle, on which the proximity between movies reflects their proximity to other movies in the user-item ratings matrix. For readability we show only the titles of the top-20 most popular movies after all 4000 movies were distributed over the available space. The three red clusters are movies that are positioned in close proximity, which we colored red and represented as a list for readability, e.g.\ a cluster with Star Wars IV and seven other movies. The distribution of the three red clusters over space indicates the existence of users that like movies in only one of these clusters. However, if we assume that there are also users that like the movies in two or even all three of these clusters, how can we construct a semantic space so that for every user an optimal region of interest exists? Using a normalized two dimensional space, there is no possible model that contains regions for all combinations of two out of three of these clusters without covering additional space. It requires a higher-dimensional space to create more overlapping regions for users with partially shared preferences. 

In a near-optimal high-dimensional semantic space, the `best' recommendation candidates are likely to be positioned in close proximity to the items the user rated highly. To recommend items to a specific user, we propose to find a function that transforms a semantic space into a one-dimensional space in which her rated items are ranked accordingly, reasoning that in the transformation the rated and unrated items that are of interest to the user will end up in a close to optimal position. 

\subsection{Related work}

A tried-and-true approach for recommending items to a user is to learn latent factors which describe the observed preferences of users towards items. Some of the most successful recommendation methods use matrix factorization to represent users and items in a shared latent low-dimensional space. The prediction of whether a user will like an item is commonly estimated by the dot product between their latent representations \cite{koren2009matrix}. The two main disadvantages to the latent factors learned are that they are not easy to interpret and that it cannot generalize beyond rated items. Different from matrix factorization, in our approach we do not optimize shared latent factors to represent users and items, but rather predict the substitutability between items. When the distance between vectors corresponds to their substitutability, the data can be interpreted more straightforward using the nearest neighbor heuristic and visualization techniques such as t-SNE. Visualization of latent factors is of interest to the recommender system community, cf. \cite{nemeth2013visualization}. We also show that both user-item ratings and textual content can be used within the same framework, which makes it possible to generalize beyond rated items, however, we leave this for future work.

Collaborative Topic Regression (CTR) fits a model in latent topic space to explain both the observed ratings and the words in a document, where the topical distribution of documents is inferred using LDA \cite{wang2011collaborative}. Dai et al.\ \cite{dai2014document} analyzed the difference between document representations that where generated by LDA and neural embeddings that were learned using the Paragraph Vector, and conclude that Paragraph Vectors significantly outperform LDA, although it is not clear why neural embeddings work better. Our model is similar to CTR in learning a model that is optimized to predict both ratings and content that is used to describe items; however, using a neural network we neither need to explicitly prescribe the type of data nor do we need to extract a topical model prior to learning the embeddings.

For item recommendation, pair-wise ranking approaches can be used the capture the pair-wise preferences over items. Baysian Personalized Ranking is a state-of-the-art approach that maximizes the likelihood of pair-wise preferences over observed and unobserved items \cite{rendle2009baysian}. However, Yao et al.\ \cite{yao2015collaborative} argue that this approach cannot incorporate additional item metadata, and is difficult to tune on sparse data. They propose to use LDA to reduce dimensionality of the data to overcome those deficiencies. In this work, we also present a pair-wise ranking approach. The key difference lies in the structure of the learned semantic space, which is learned with a Paragraph Vector architecture, chosen with the goal of making regions of interest more easily separable when dealing with a large number of dimensions. In a sense, such a space resembles a metric space, meaning that our approach can be viewed as a proposal to learn a ranking function based on vector algebra rather than by estimated likelihood.

For the task of recommending movies, Musto et al.\ \cite{musto2016learning} use semantic vectors for movies that are the average over the Word2Vec embeddings of the words on the movie's Wikepedia page. In our approach the semantic vectors are learned to jointly predict observations for movies, rather than an average over the semantic vectors of individual words. For the recommendations, Musto et al.\ regard a user's preference as the average vector of their highly rated movies, and then movies are ranked according to their distance to this point in semantic space. In this work, instead of positioning the user in semantic space a function is learned that transforms the structure in semantic space into a ranking that is optimized for a user's past preferences.

For the task of personalizing relevant text content to users, Elkahky et al.\ \cite{elkahky2015multi} propose a content-based approach to map users and items to a shared semantic space, and recommend items that have maximum similarity to a user in the mapped space. By jointly learning a space using features from clicked webpages, news articles, downloaded apps and viewed movie and TV program, they show that recommendations improve over those only learned over a single domain. Following the Deep Structured Semantic Model (DSSM) that was proposed in \cite{huang2013learning}, user and item features are mapped to 128-dimensional semantic vectors using a 5-layer architecture to maximizing the similarity between the semantic vectors of users and the items they interacted with in the past. In our work, a shallow neural network is used to learn item vectors that optimally predict their observed features using a shared weight matrix. To recommend items for a single user, the user-independent space is transformed according to their past preferences.


\section{Approach}
\label{section:design}

\subsection{Learning semantic vectors}

Bengio et al.\ \cite{bengio2003neural} propose to learn embeddings for words based on their surrounding words in natural language. Although the architecture that Bengio et al.\ proposed is still applicable for learning state-of-the-art semantic vectors, their approach received only moderate attention until Mikolov et al.\ \cite{mikolov2013distributed} used this idea to design highly efficient deep learning architectures for learning embeddings for words and short phrases, also known as Word2Vec. They show that the accuracy of the word embeddings increases with the amount of training data, and to some extent that the learning process consistently encodes some generalizations in the semantic vectors which can be used for analogous reasoning, such as the gender difference between otherwise equivalent words. This generalizing effect possibly occurs when a more efficient encoding can be used to jointly predict similar contexts for different words, although the exact conditions under which these generalizations are captured are not known. Recently, Le and Mikolov \cite{le2014distributed} proposed an architecture to learn embeddings for paragraphs and documents.

\begin{figure}[!tbp]
  \centering
\begin{tikzpicture}
   \foreach \i in {-3, ..., 3}{
		\draw (\i / 2 - 0.25,4) rectangle (\i / 2 + .25,4.5);
	}
   \node at (-1.5, 4.25) { $y_0$ };	
   \node at (-1, 4.25) { $y_1$ };	
   \foreach \i in {-1, ..., 2}{
      \node at (\i / 2, 4.25) { $...$ };	
   }	
   \node at (1.5, 4.25) { $y_m$ };	
   \foreach \i in {-4, ..., 4}{
		\draw (\i / 2 - 0.25,2) rectangle (\i / 2 + .25,2.5);
		\foreach \j in {-3, ..., 3}{
		   \draw[] (\i / 2, 2.5) -> (\j / 2, 4);
		   \draw[] (\i / 2, 2) -> (\j / 2, .5);
		}
	}
   \node at (-2, 2.25) { $h_0$ };	
   \node at (-1.5, 2.25) { $h_1$ };	
   \foreach \i in {-2, ..., 3}{
      \node at (\i / 2, 2.25) { $...$ };
   } 
   \node at (2, 2.25) { $h_i$ };	
   \foreach \i in {-3, ..., 3}{
		\draw (\i / 2 - 0.25,0) rectangle (\i / 2 + .25,.5);
	}
   \node at (-1, .25) { $m_1$ };	
   \node at (-1.5, .25) { $m_0$ };	
   \foreach \i in {-1, ..., 2}{
      \node at (\i / 2, .25) { $...$ };
   } 
   \node at (1.5, .25) { $m_n$ };	
	
   \node[fill=white] at (0,1.25) { ($w_0$) };
   \node[fill=white] at (0,3.25) { ($w_1$) };
   \draw[dashed] (-2, 3.8) to (2, 3.8);
   \node at (2.5, 3.8) { sigmoid };

\end{tikzpicture}
   \caption{Deep learning architecture that is used to learn semantic vectors for items. The observations are streamed one-at-a-time, placing a movie-id in the input layer (bottom), which lookup a embedding in $w_0$ and places it in the hidden layer (middle). The model then updates weights $w_1$ of the observed item $y$ and the embedding to optimize predictions using stochastic gradient descent.}
  \label{fig:skipgram}
\end{figure}
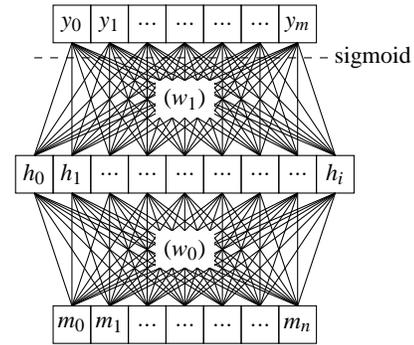

In this study, semantic vectors for the items in a corpus are learned using the \emph{Paragraph Vector} architecture described in Figure~\ref{fig:skipgram}, which is similar to the PV-DBOW architecture proposed by Le and Mikolov \cite{le2014distributed}. The input (bottom) is a `1-hot lookup' vector, that contains as many nodes as there are items, and for every training sample only has the node that corresponds to the movie ID set to 1 while the other nodes are set to zero, which effectively looks up an embedding for a given movie $m$ in weight matrix $w_0$ and places it in the hidden layer (middle). The output layer contains a node $y$ for every possible observation in the training samples. The weight matrices $w_0$ and $w_1$ respectively connect all possible input nodes, hidden nodes and output nodes. We learn the embeddings by predicting the outputs in a hierarchical softmax, i.e.\ all possible outputs are placed in a binary Huffman tree to learn the position of the observation in the tree rather than separate probabilities for each possible output \cite{mikolov2013distributed}. The item embeddings are learned together with a weight matrix $w_1$ by streaming over the observed features one-at-a-time in random order. For every movie, the network can generate a probability distribution over all possible observations by computing the dot product between the embedding with $w_1$. Using stochastic gradient descent, the embeddings and weights are updated to improve the prediction of the observed data. The learning process is similar to that described by Mikolov et al.\ \cite{mikolov2013distributed} for the learning of word distribution using a Skipgram architecture against a hierarchical softmax, except that no context window is used but rather all observations are processed one-at-a-time.

To learn semantic vectors that capture the substitutability between items, the observations used to learn the semantic vectors should be representative for their substitutability. This can for instance be inferred from the observation that a group of users gave these items high ratings, but also from reviews that each describe an item or an opinion about the item. Lops et al.\ \cite{lops2011content} argue that existing content-based techniques require knowledge of the domain, however, learning item representations using a neural network has the advantage that patterns between items are learned automatically and therefore obviates the need for prior domain knowledge. In the evaluation, we will show that we can effectively learn semantic vectors for items using the same deep learning architecture on both user-item ratings as well as item contents. 

In this study, we preprocessed the data for use with the Paragraph Vector. To correct for the anchoring effects mentioned in \cite{koren2010collaborative}, the ratings are interpreted as relative to its user's average, replace ratings below the user's average with a rating of 1 and equal or above the average with a rating of 2. These semantic vectors are learned from paired training samples $(\mathit{item\ ID}, \mathit{observation})$, where the observation can be an attribute of an item, a word that appear in an item's description (in this study a movie review), or an item's rating by a user. The input is transformed so that every observation becomes a single word, e.g. for Star Wars IV, which has id 240 in MovieLens 1M the rating 3 by user 73 (who has given an average rating of 3.4) is transformed into (240, `user73\_rating1') and in a content-based setting a review fragment that contains ``The masterpiece, the legend that made people...'' is transformed into (240, `the'), (240, `masterpiece'), (240, `the'), etc.. 

\subsection{User-specific ranking}

In Section~\ref{sec:theorydeepranking}, we argued that for a near-optimal semantic space there should be a function that transforms this semantic space into a one-dimensional space in which a user's past preferences lie according to their ratings. In this work, we limited our search for such a function to finding a hyperplane for this transformation. Such a hyperplane is described by a normalized vector that is orthogonal to the hyperplane, and the dot product with this vector projects the semantic vectors to a one-dimensional space according to their squared distance to the hyperplane, which is negative for items that lie on the opposite side of the hyperplane. By using a hyperplane, dimensions that are less useful for ranking the items can be down weighted or even ignored by choosing a hyperplane parallel to those dimensions. 

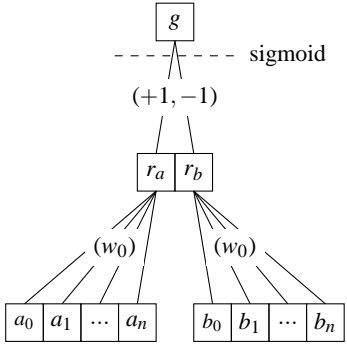
\begin{figure}[!tbp]
  \centering
\begin{tikzpicture}
   \foreach \i in {0}{
		\draw (\i / 2 - 0.25,4) rectangle (\i / 2 + .25,4.5);
	}
   \node at (0, 4.25) { $g$ };	
   \foreach \i in {-.5, .5}{
		\draw (\i / 2 - 0.25,2) rectangle (\i / 2 + .25,2.5);
		\foreach \j in {0}{
		   \draw[] (\i / 2, 2.5) -> (\j / 2, 4);
		}
	}
   \node at (-.25, 2.25) { $r_a$ };	
   \node at (+.25, 2.25) { $r_b$ };	
   	
   \foreach \i in {1, ..., 4}{
		\draw (\i / 2 - 0.25,0) rectangle (\i / 2 + .25,.5);
		\draw (-\i / 2 - 0.25,0) rectangle (-\i / 2 + .25,.5);
		\draw (\i / 2, .5) to (.25, 2);
		\draw (-\i / 2, .5) to (-.25, 2);
	}
   \node at (-2, .25) { \small $a_0$ };	
   \node at (-1.5, .25) { $a_1$ };	
   \node at (-1, .25) { $...$ };
   \node at (-.5, .25) { $a_n$ };	
	
   \node at (.5, .25) { \small $b_0$ };	
   \node at (1, .25) { $b_1$ };	
   \node at (1.5, .25) { $...$ };
   \node at (2, .25) { $b_n$ };	
	
   \node[fill=white] at (-.8,1.25) { ($w_0$) };
   \node[fill=white] at (.8,1.25) { ($w_0$) };
   \node[fill=white] at (0,3.25) { ($+1, -1$) };
   \draw[dashed] (-.8, 3.8) to (.8, 3.8);
   \node at (1.5, 3.8) { sigmoid };

\end{tikzpicture}
   \caption{Neural network architecture that is used to learn the parameters $w_0$ of a hyperplane that optimally transforms items from an n-dimensional semantic space into a one-dimensional space, by optimizing the predicted order of pairs of item vectors $a$ and $b$ as rated by a user. The item pairs are streamed one-at-a-time, placing the semantic vector of the lower rated item in $a$ and of the higher rated item in $b$. Starting with a random hyperplane $w_0$ the scores $r_a, r_b$ are computed and the resulting gradient $g$ is used to rotate the hyperplane towards a more optimal ranking using stochastic gradient descent.}
  \label{fig:deeplearning}
\end{figure}

To learn an optimal hyperplane, we propose a neural network architecture that optimizes the ranking over pair-wise preferences. Figure~\ref{fig:deeplearning} shows a schematic of the architecture, which learns a hyperplane orthogonal to $w_0$ by stochastic gradient descent over pairs of item vectors $a$ and $b$, given that item $a$ has received a lower rating than $b$. The semantic vectors for $a$ and $b$ are not updated during learning. A shared weight matrix $w_0$ is used to compute a score of respectively $r_a$ and $r_b$ as the dot product between the semantic vectors and $w_0$. These scores are then combined using the fixed weights $(+1, -1)$, and filtered by a sigmoid function. The output layer directly provides the gradient $g \in [0,1]$ that is used to update $w_0$, by subtracting $g \cdot \alpha \cdot a$ from $w_0$ and adding $g \cdot \alpha \cdot b$ to $w_0$. The learning rate $\alpha$ linearly descends from an initial value (in this study by default 0.025) to 0 during the learning process.

When estimating an optimal hyperplane to transform a semantic space into a ranking, all unrated items are considered to be 0. Similar to the preprocessing used for learning the semantic vectors, ratings are replaced by 1 if they are below the user's average and with 2 if they are equal or above the user's average, to correct for anchoring effects \cite{koren2010collaborative}. When learning the hyperplane, the system iterates $\phi_i$-times over all item pairs that are rated differently by the user. 

The time needed to learn the parameters of a hyperplane increases quadratically over the number of items the user has rated. Interestingly, there are several way to improve both the efficiency and effectiveness of the learning process. Koren observed that users' preferences change over time and shift between concepts \cite{koren2010collaborative}. We hypothesize that simply using only the $\phi_t$-most recently rated items may improve both the effectiveness and the efficiency of the recommender system. Another consideration for item recommendation is that optimally predicting the higher ranked items is more important than the ranking between lower ranked items. Typically, relatively few of the available items are of interest to the average user, and to avoid over-optimizing the prediction of unrated items over interesting items the unrated items can be down sampled. In this work, the down-sampling rate is controlled by a hyperparameter $\phi_d$, e.g.\ when $\phi_i=10$ iterations are used with downsampling $\phi_d=0.1$ every combination between a rated and an unrated item is used in exactly one randomly chosen iteration, while the combinations between two rated items are used $\phi_i$ times for learning.


\section{Experiment}
\label{section:experiment}

The proposed ``Deep Space'' approach (DS) first learns user-inde-pendent semantic vectors for items, which can then be transformed into a ranking that is optimized according to a single user's preferences. We will show that by using only the $\phi_t$ items the user rated prior to the time of recommendation, both efficiency and effectiveness are greatly improved. However, in order to have a timestamp to determine the most recent ratings the evaluation should use a leave-one-out evaluation strategy. Since our semantic space model is currently not-updatable, using a leave-one-out strategy on the entire dataset is not feasible since for every item a new semantic space must be learned. To implement a fair, yet feasible, test procedure, we sampled a test-set from the dataset that consists of a user's temporarily latest ratings, then a single semantic space is learned using all ratings except those in the test set, and in the evaluation this model is used to predict the test samples. For this reason, the experimental systems use no information that lies in the future with respect to the target user at the moment of interaction with the test item.

In this paper, we carry out initial experiments that test the viability of the ``Deep Space'' approach. We chose MovieLens 1M because it is easily available and its properties are well-known, making it easy for others to understand and reproduce our findings. Note that we need a data set in which both ratings and reviews are available for the items. The MovieLens 1M dataset consists of 1 million ratings by 3952 users for 6040 movies on a 5-point scale. For the content-based experiments, we use the contents of the movies' user reviews on IMDB without their rating or username, and consider every word in the review text an observed word. To sample a validation and test set, we order the users by their number of ratings, and the ratings by the time they were submitted. Then, in that order of all ratings by all users, we mark every 25th rating. This ensures the test set matches the corpus' distribution over users rating volume, since prediction difficulty may be different between users that rated a few or many items. Then, if for a user $n$ ratings are marked, from her temporarily-last $n$ ratings the first half is assigned to the validation set, the last half to the test set, and in case of an odd number it is assigned to the shorter of the two sets or the validation set when equal in length. The models' parameters are tuned using the validation set, by training the model on all ratings except those in the test or the validation set. For the evaluation we use the test set after training the models on all ratings except those in the test set. All systems use the exact same training, validation and test set for the evaluation.

The effectiveness of the recommender systems is evaluated using Recall@10 over the approximately 10k ratings in the test set that are a 4 or a 5 on a 5-point scale. The Recall@10 metric is directly interpretable as the proportion of left-out items that a system returns in the top-10 recommendations.


\section{Results}
\label{section:results}

We evaluate the effectiveness of our approach, by comparing the results of our approach to that of a popularity baseline and the MyMediaLite implementation of BPRMF \cite{rendle2009baysian}, WRMF \cite{hu2008collaborative}, and UserKNN. The parameters for all models are tuned on the validation set that is described in Section~\ref{section:experiment}, and the resulting parameters are shown in Table~\ref{table:parameters}. 
\begin{table}
\centering
\caption{Parameters tuned for MovieLens 1M}
\label{table:parameters}
\begin{tabular}{|l|l l|}
\hline
System   & Recall@10 & \\
\hline
BPRMF	&  $factors=100$, $reg=0.001$, $lrate=0.025$, $iter=30$ &\\
WMRF	&  $factors=20$, $reg=0.020$, $alpha=0.1$, $iter=10$ &\\
UserKNN	&  $k=60$ & \\
\hline
DS-CB	& $\phi_d = 1$, $\phi_t = 10$, $\phi_i = 10 $ & \\
DS-VSM	& $\phi_d = 20$, $\phi_t = 5$, $\phi_i = 10$ & \\
DS-CF	& $\phi_d = 20$, $\phi_t = 5$, $\phi_i = 10$ & \\
\hline
\end{tabular}
\end{table}

For the proposed model, we evaluate three variants: The DS-CB variant uses the Paragraph Vector to learn semantic vectors from the text of IMDB user reviews, and uses no rating information of other users than the user that is recommended to. The DS-VSM variant does not learn a contiguous semantic space using the Paragraph Vector, but uses a normalized vector space model (VSM) in which every user is a dimension and each item is represented as a vector consisting of its user ratings. The DS-CF variant uses the Paragraph Vector to learn a semantic space from the user-item ratings from which the recommendations are made. Table~\ref{table:recall} reports Recall@10 obtained by all models on the test set. We tested the differences between systems for statistical significance, using the McNemar test on a 2x2 contingency table of paired nominal results (a left-out item is retrieved in the top-10 of neither, one or both systems). In Table~\ref{table:recall}, all significant improvements have a $\textit{p-value} < 0.001$. In these experiments, the DS-CF and DS-VSM models are significantly more effective than BPRMF, WRMF, UserKNN and DS-CB. By including the DS-VSM model in the evaluation, we show that the improvement is not only the result of learning semantic vectors with the Paragraph Vector, but is partially contributed by learning a hyperplane to optimally rank a user's past ratings for the recommendation. However, since the DS-CF variant significantly outperforms the DS-VSM variant, we also show the benefit of learning semantic vectors with the Paragraph Vector which for generating recommendations is both more effective and more efficient. Although the representations learned with the Paragraph Vector are lower in dimensionality than the VSM over all users, typically, the DS-CF performs best in much higher-dimensional space than state-of-the-art matrix factorization approaches. The DS-CB variant that learns 10k dimensional semantic vectors from movie reviews is significantly less effective than the approaches that use user-item ratings. However, for items that have not been rated the content-based variant may provide an alternative. 

\begin{table}
\centering
\caption{Comparison of the effectiveness on MovieLens 1M. The subscripts in the column ``sig. over'' correspond to a significant improvement over the corresponding system, tested using McNemar test, 1-tailed, $\textit{p-value} < 0.001$.}
\label{table:recall}
\begin{tabular}{|l|r l|}
\hline
System   & Recall@10 & sig. over \\
\hline
Pop 		& 0.053 &\\
BPRMF $^1$	& 0.079& $^4$ \\
UserKNN $^2$ & 0.087 & $^4$ \\
WMRF $^3$	& 0.089 & $^4$ \\
\hline
DS-CB-10k $^4$	& 0.075 & \\
DS-VSM $^5$	& 0.119 & $^{1,2,3,4}$ \\
DS-CF-500	& 0.144 & $^{1,2,3,4,5}$ \\
DS-CF-1k		& 0.151 & $^{1,2,3,4,5}$ \\
\hline
\end{tabular}
\end{table}

\begin{figure*}[t]
   \centering
   \begin{subfigure}[b]{.3\textwidth}
       \includegraphics[width=0.9\textwidth]{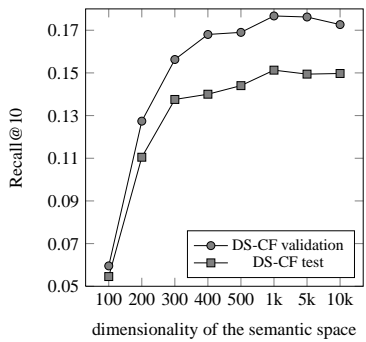}
       \caption{The effect that the dimensionality has on effectiveness.}
       \label{fig:dimensions}
   \end{subfigure}
   \qquad
   \begin{subfigure}[b]{.3\textwidth}
       \includegraphics[width=0.9\textwidth]{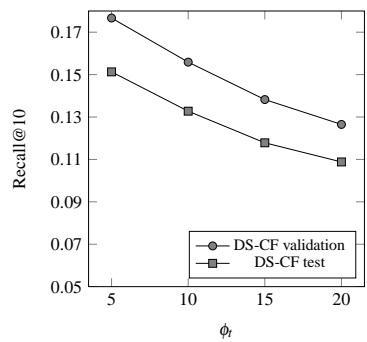}
       \caption{The effect that using only the number most recently rated movies has on effectiveness.}
       \label{fig:trim}
   \end{subfigure}
   \qquad
   \begin{subfigure}[b]{.3\textwidth}
       \includegraphics[width=0.9\textwidth]{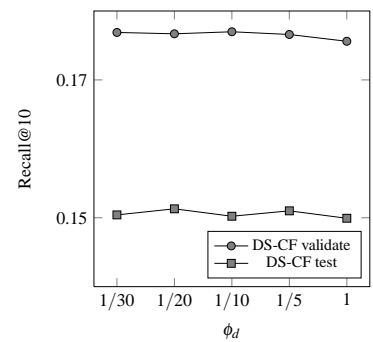}
       \caption{The effect that downsampling the use of unrated items has on effectiveness.}
       \label{fig:downsampling}
   \end{subfigure}
   \caption{Sensitivity of hyperparameters}
\end{figure*}

We analyze the sensitivity of the hyperparameters $\phi_d$, $\phi_t$ and the dimensionality of the semantic space. Hereto we perform a sweep over these parameters using the DS-CF model, changing only one hyperparameter at a time while setting the remaining two out of three parameters to $dimensionality=1000$, $\phi_t=5$, and $\phi_d=20$. In Figure~\ref{fig:dimensions}, by changing the dimensionality of the semantic space we observe that the DS-CF model outperforms the VSM variant when dimensionality is at least 300, and that the effectiveness does not improve beyond the use of 1k dimensions. The degradation in performance when using less than 300 dimensions is possibly related to the linear transformation function that is used to rank the items, since in a lower dimensional space it may not be possible to position the items so that for all users there exists a linear function to generate a close to optimal ranking. In Figure~\ref{fig:trim}, we observe that using only the $n$ most-recent ratings given by a user is more effective for lower values of $n$; when using more than five ratings to learn a transformation function the effectiveness degrades. In Figure~\ref{fig:downsampling}, shows the effect that down sampling of the used unrated items has on the effectiveness of learned transformation functions, where $\phi_d = 1$ equals no down sampling. In general, down sampling improves the efficiency of the recommendation while not having any negative impact on recall. This hyperparameter does not appear to be sensitive on this collection. The optimal value for these three hyperparameters may be collection dependent, and therefore need to be tuned.

\begin{figure}[H]
   \centering
       \includegraphics[width=0.4\textwidth]{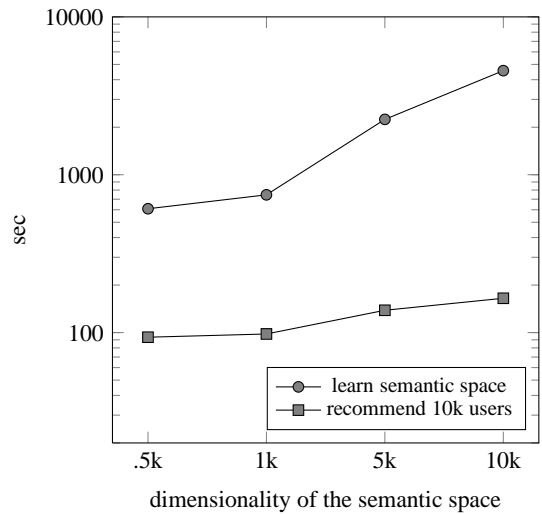}
       \caption{The time to learn a semantic space using the Paragraph Vector on user-item ratings and the time to generate 10k recommendations by hyperplane projection.}
       \label{fig:efficiency}
\end{figure}

We finally report about the efficiency of the proposed approach. All experiments were performed on a machine with two Intel(R) Xeon(R) CPU E5-2698 v3, which together have 32 physical cores. Using the test-set as described in Section~\ref{section:experiment}, Figure~\ref{fig:efficiency} reports the wall time in seconds for learning a semantic space with the Paragraph Vector on the user-item ratings on the training data of the test-set, and the total time taken to generate a full ranked list for the approximately 10,000 items in the test set. For learning the semantic spaces, the user-item ratings were processed in 20 iterations, which for a 1000 dimensional semantic space takes 12.5 minutes. For the same dimensionality, the average time to rank all items using the parameter settings in Table \ref{table:parameters} according to a user's preferences takes approximately 0.3 core seconds.

\section{Conclusion}

For the task of recommending items to a user, we propose to learn a semantic space in which substitutable items are positioned in close proximity. We show that these spaces can be learned from item reviews as well as user-item ratings, using the same deep learning architecture. To recommend items to a specific user, we learn a function that optimally transforms a user-independent semantic space into a ranking that is optimized according to the user's past ratings. In the experiments that use user-item ratings, this approach significantly outperformed BPRMF, WRMF and UserKNN on the MovieLens 1M dataset. When a semantic space is learned from user reviews on IMDB, the results are not as effective as these existing collaborative-filtering baselines, but may be useful to recommend novel items or when there is an insufficient amount of user-item ratings available to use collaborative filtering.

An interesting direction for future work is to extend function space to non-linear functions, that are potentially more optimal when the dimensionality of the semantic space is reduced. Another interesting direction is to jointly learn item representation based on content and collaborative filtering data, which may improve recommendation on sparse collections and for cold start cases.

\section*{Acknowledgment}

This work was carried out on the Dutch national e-infrastructure with the support of SURF Foundation. The second author is partially funded by EU FP7 project CrowdRec (610594).

\bibliographystyle{abbrv}
\bibliography{bibliography}  

\end{document}